\documentclass[lettersize,journal]{IEEEtran}

\usepackage[colorlinks,urlcolor=blue,linkcolor=blue,citecolor=blue]{hyperref}
\expandafter\def\expandafter\UrlBreaks\expandafter{\UrlBreaks\do\/\do\*\do\-\do\~\do\'\do\"\do\-}

\usepackage{wrapfig}
\usepackage{hyperref}
\usepackage{amsmath}
\usepackage{amsthm}
\usepackage{graphicx}
\usepackage{textcomp}

\usepackage{comment}
\usepackage{tabularx,tabulary}
\usepackage{tabularx,ragged2e,booktabs}
\usepackage[skip=1pt]{caption}

\usepackage{etex}
\usepackage{algorithmicx}
\usepackage{algpseudocode}
\usepackage[ruled,vlined]{algorithm2e}
\SetKwRepeat{Do}{do}{while}

\usepackage[utf8]{inputenc}
\usepackage{multicol}
\usepackage{multirow}
\usepackage{comment}
\usepackage{colortbl}
\usepackage{pstricks}

\usepackage{pifont}
\usepackage{framed}
\usepackage{multirow}
\usepackage{array}
\newcolumntype{L}[1]{>{\raggedright\let\newline\\\arraybackslash\hspace{0pt}}m{#1}}
\newcolumntype{C}[1]{>{\centering\let\newline\\\arraybackslash\hspace{0pt}}m{#1}}
\newcolumntype{R}[1]{>{\raggedleft\let\newline\\\arraybackslash\hspace{0pt}}m{#1}}
\usepackage{url}

\newcommand{\newcontent}[1]{{#1}}

\usepackage[most]{tcolorbox}

\usepackage{booktabs}

\usepackage[noadjust]{cite} 
\usepackage{setspace}
\usepackage{amsmath}
\usepackage{amssymb} 
\usepackage{amsfonts} %
\usepackage{calligra} %
\usepackage{mathtools}
\usepackage{amsthm} 
\usepackage{array}
\usepackage{subcaption}

\usepackage{pgfplots}
\usepackage{pgfplotstable}
\usepgfplotslibrary{statistics}

\usepackage{comment}
\usepackage{listings}

\usepackage{circledsteps}

\usetikzlibrary{patterns}

\usepackage{acronym}

\input{preamble/preamblelistingsconf}
\acrodef{CPS}{Cyber-Physical System}
\acrodef{IoT}{Internet of Things}
\acrodef{HDL}{Hardware Description Language}
\acrodef{CAD}{Computer-Aided Design}
\acrodef{EDA}{Electronic Design Automation}
\acrodef{HPC}{High-Performance Computing}
\acrodef{DL}{deep learning}
\acrodef{ML}{machine learning}
\acrodef{NLP}{natural language processing}
\acrodef{IC}{Integrated Circuit}
\acrodef{CWE}[CWE]{Common Weakness Enumeration}
\acrodef{CVE}[CVE]{Common Vulnerabilities and Exposures}
\acrodef{LLM}[LLM]{large language model}
\acrodef{NMT}[NMT]{neural machine translation}
\newcommand{\ignore}[1]{{}}

\newcommand{\squishlist}{
	\begin{list}{$\bullet$}
		{ \setlength{\itemsep}{0pt}
			\setlength{\parsep}{1pt}
			\setlength{\topsep}{1pt}
			\setlength{\partopsep}{0pt}
			\setlength{\leftmargin}{0.9em}
			\setlength{\labelwidth}{1.5em}
			\setlength{\labelsep}{0.4em} } }
	\newcommand{\squishend}{
	\end{list}  }

\definecolor{graphFirst}{RGB}{2,136,209} 
\definecolor{graphSecond}{RGB}{211,47,47} 
\definecolor{graphThird}{RGB}{245,124,0} 
\definecolor{graphFourth}{RGB}{56,142,60} 
\definecolor{graphFifth}{RGB}{81,45,168} 
\definecolor{graphSixth}{RGB}{69,90,100} 
\definecolor{graphSeventh}{RGB}{251,192,45} 
\definecolor{backgroundSecond}{RGB}{239,154,154} 
\definecolor{backgroundThird}{RGB}{255,204,128} 
\definecolor{backgroundFourth}{RGB}{165,214,167} 
\definecolor{backgroundFifth}{RGB}{179,157,219} 
\definecolor{backgroundSixth}{RGB}{176,190,197} 
\definecolor{backgroundSeventh}{RGB}{255,245,157} 

\begin{document}
\bstctlcite{IEEEexample:BSTcontrol}

\title{Enabling Deep Visibility into VxWorks-Based Embedded Controllers in
  Cyber-Physical Systems for Anomaly Detection}

\author{Prashanth Krishnamurthy, Ramesh Karri, and Farshad Khorrami
        \thanks{P. Krishnamurthy, R. Karri, and F. Khorrami are with the
                Dept. of ECE, NYU Tandon School of Engineering, Brooklyn, NY 11201, USA.
                (e-mails: \{prashanth.krishnamurthy, rkarri, khorrami\}@nyu.edu).}
        \thanks{
        This work was supported in part by
NSF SaTC grant 2039615, DOE NETL contract DE-CR0000017, and
DARPA under AFRL contract FA8750-16-C-0179.
        The views and conclusions contained in this document are those of the authors and should not be interpreted as representing the official policies, either expressed or implied, of the U.S. Government.
        }
        }

\maketitle

\begin{abstract}
  We propose the DIVER (Defensive Implant for Visibility into Embedded
  Run-times) framework for real-time deep visibility into embedded control
  devices in cyber-physical systems (CPSs). DIVER enables run-time detection of
  anomalies and targets devices running VxWorks real-time operating system
  (RTOS), precluding traditional methods of implementing dynamic
  monitors using OS (e.g., Linux, Windows) functions. DIVER has two components:
  ``measurer'' implant embedded into VxWorks kernel to collect
  run-time measurements and provide interactive/streaming interfaces over
TCP/IP; remote ``listener'' that acquires and analyzes measurements
  and provides interactive user interface. DIVER focuses on small embedded
  devices with stringent resource constraints (e.g., insufficient storage to
  locally store measurements). \newcontent{To show efficacy and scalability of DIVER, we
  demonstrate on two embedded devices with different processor architectures
  and VxWorks versions: Motorola ACE Remote Terminal Unit used in CPS including power systems and Raspberry Pi representative of Internet-of-Things (IoT) applications.}
\end{abstract}

\begin{IEEEkeywords}
Anomaly detection, PLC, real-time operating system, VxWorks, embedded systems, defensive implant, remote monitoring.
\end{IEEEkeywords}

\section{Introduction}
\label{sec:introduction} With growing complexity, connectivity, and remote
programmability and configurability of embedded control devices in
CPSs, robust cyber-security and anomaly detection techniques
are becoming increasingly vital~\cite{KKK_dtm2016,SBS_plc_blaster_2016,urgent11}. The need for such techniques is becoming more
crucial with sophisticated adversaries able to transit from the information
technology (IT) network to the operational technology (OT) network and exploit
vulnerabilities of embedded devices to insert malicious code, alter device
configurations, or modify their behavior. Through such adversarial
manipulations, adversaries can severely disrupt the CPS operation and cause catastrophic consequences to
 CPS stability, safety, or performance. Hence, development of
real-time monitoring and anomaly detection techniques has been intensely studied
and several approaches have been proposed using techniques such as network
traffic monitoring, host-based intrusion detection, side channel analysis, etc.

While host-based methods using operating system (OS) level observations such as
system call traces and registry modifications can be effective for devices
running general-purpose OS such as Linux or Windows, they are not
directly applicable to embedded devices that run real-time operating systems
(RTOS) such as VxWorks, QNX, or FreeRTOS. Such embedded devices are prevalent in
CPS such as power grid due to their real-time performance, robustness,
simplicity, and reliability. The development of corresponding host-based
monitoring methods for such devices faces multiple challenges due to, for
example, monolithic structure of such RTOS (e.g., no separate processes),
software ecosystem constraints (e.g., do not expose command-line
shell or only provide a shell under special debug mode that requires device
reset and is very different from normal mode, difficulties in deploying custom
code), and device-level limitations (e.g., limited disk space and computing resources). While human-machine interfaces (HMIs) of these devices
provide some visibility into device operation, on-device malware can
 spoof these observations and cloak their presence. Hence,
deep under-the-hood visibility into device run-time operation is crucial to
establish integrity of the devices.

\begin{figure*}[!h]
  \centering
  \includegraphics[width=6in]{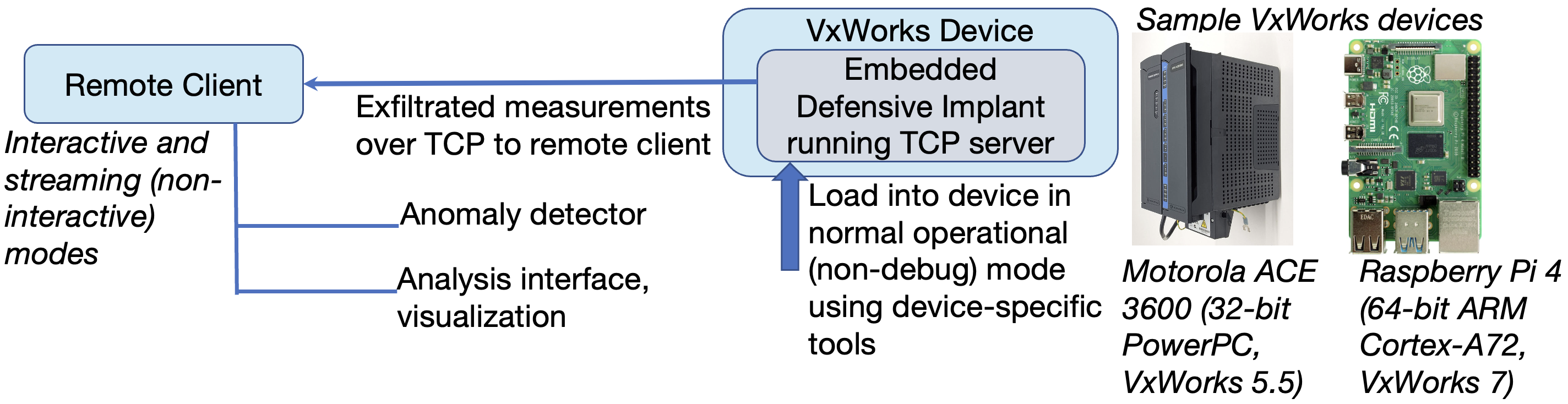}
  \hrule
  \includegraphics[width=\textwidth]{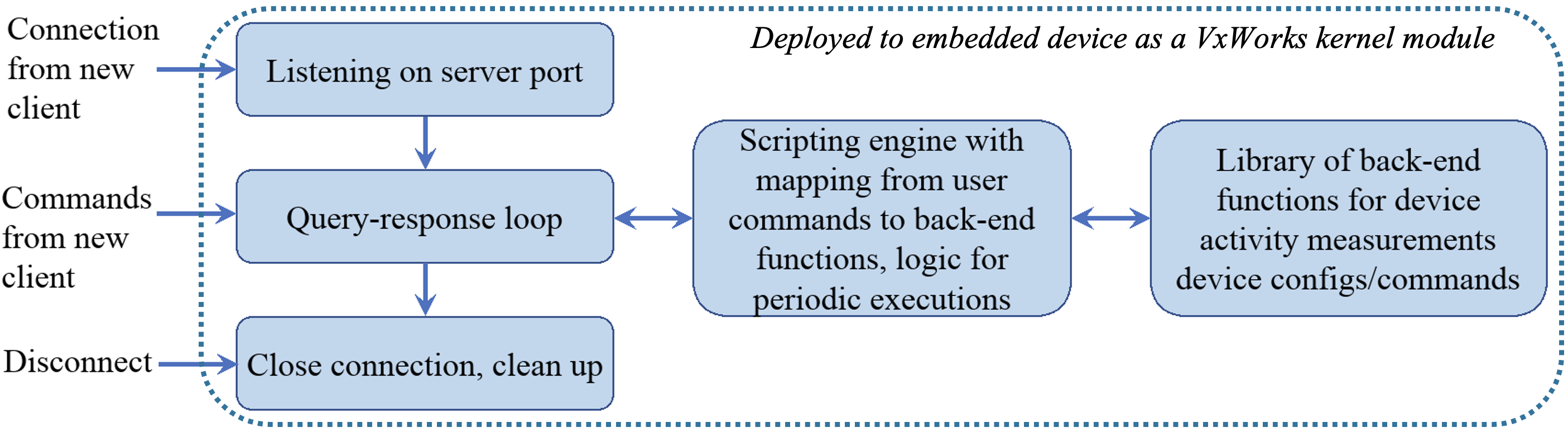}
  \vspace*{-0.05in}
  \hrule
  \vspace*{0.08in}
  \includegraphics[width=0.8\textwidth]{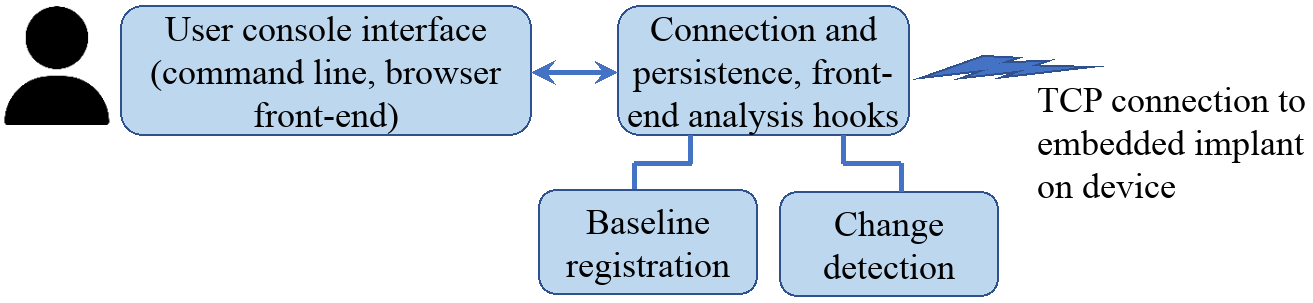}
  \caption{Structure of DIVER framework along with sample devices on which DIVER has been tested (top). Architectures of on-device defensive implant or ``measurer'' (middle) and off-device remote client or ``listener'' (bottom).}
  \label{fig:block_diagram}
\end{figure*}

To address this vital need, we develop the DIVER framework (Figure~\ref{fig:block_diagram}) for achieving
real-time deep visibility into embedded devices (e.g., programmable logic controllers -- PLCs, remote terminal units -- RTUs). DIVER enables real-time introspection and anomaly detection by embedding a
defensive implant (``measurer'') directly into the RTOS to enable under-the-hood
monitoring of run-time measurements (of various types including task-level activity measurements, device status, firmware module information, timer interrupt configurations, memory and filesystem contents, etc.). This implant communicates with a remote monitoring agent
(``listener'') via TCP/IP to exfiltrate real-time measurements from
within the RTOS. The listener analyzes these measurements to build a real-time
situational awareness of the device execution state and detect anomalies
relative to prior observations or baselines. The listener also provides an
interface for interactive commands to be relayed to the VxWorks-embedded implant
-- this interface can be used both by human operators or automated scripts.
\newcontent{DIVER's main focus is to enable deep real-time visibility
into VxWorks-based embedded devices with stringent resource constraints and the
multiple challenges discussed above. DIVER enables
lightweight RTOS-integrated measurement and interactive
mechanisms in the embedded device and real-time exfiltration to the remote
listener. It is noted that the exfiltrated time series measurements (which
also include on-demand exfiltration of memory ranges from the device) enabled by
DIVER can
then be utilized along with the wide range of binary analysis and anomaly
detection algorithms available in the literature.}
DIVER focuses on small embedded resource-constrained devices such as
PLCs and RTUs in CPS.
\newcontent{For proof-of-concept and to demonstrate generality and scalability of DIVER, we demonstrate the framework on two different embedded devices with different processor architectures and VxWorks versions: Motorola ACE RTU~\cite{motorola_ace} with PowerPC processor and
running VxWorks 5.5, Raspberry Pi 4 device with ARM Cortex processor and running VxWorks 7.}

\section{Background and Related Work}
\label{sec:background} While RTOS such as VxWorks and QNX are popularly used in
embedded devices due to their deterministic real-time properties,
performance, robustness, and reliability, they are not immune to vulnerabilities
that can be exploited by adversaries. For example, zero-day
vulnerabilities (URGENT/11) discovered in VxWorks in
2019~\cite{urgent11} were found to potentially affect billions of devices across
various industries. Vulnerabilities specific to particular devices such as
Motorola ACE RTU have also been noted~\cite{motorola_ace_vulnerability}. Several
classes of vulnerabilities have been identified across a broad range of embedded
devices across several major vendors~\cite{WSG_insecure_2023} and
found to potentially allow attackers to gain unauthorized access to
manipulate configurations, modify code, and alter behavior.
These observations underscore need for robust real-time
monitoring and anomaly detection. Devices running general-purpose OS such as
Linux and Windows can leverage OS-level functionalities~\cite{KSDKK_iccd2019} to
facilitate monitoring and anomaly detection using various ``side channel''
observations (e.g., system call traces, stack traces, registry keys,
enumerations of processes and dynamically loaded modules, Hardware Performance
Counters -- HPCs, etc.). On the other hand, devices running RTOS such as VxWorks
pose unique challenges to development of corresponding monitoring methods. While
network-based monitoring methods (such as protocol payload anomaly detection~\cite{KKJKS_unknown_2021}) can be device-agnostic, host-based methods need
to address challenges of integrating into an RTOS, deploying such custom
code, extracting run-time measurements for a comprehensive view of
device operation, and exfiltrating measurements to remote monitoring
agents. These challenges have also been discussed in
\cite{MV_host_intrusion_2021}, where a host intrusion detection system was
developed to collect system logs and RAM usage measurements from a Bosch Rexroth
AG IndraControl XM device with Intel Atom processor running VxWorks.
Since integrating monitoring methods into device RTOS is challenging,
off-device approaches have been studied via methods such as evaluating input-output behaviors using offline copies of
PLC programs~\cite{CPS_code_integrity_2021}, symbolic execution and model
checking to analyze PLC control logic~\cite{ZRM_detecting_2014}, and
measurements of program execution times using vendor-provided tools to detect
timing anomalies during PLC scan cycles~\cite{FB_temporal_2020}.

In comparison with prior work discussed above, the DIVER framework provides
several unique contributions \newcontent{with a specific focus on resource-constrained
VxWorks-based embedded devices}: (1) it enables dynamic task-level
measurements providing deep visibility into run-time behavior rather than
relying upon built-in logging which could be circumvented by
adversaries -- the dynamic time series measurements are analogous to
process-level measurements that could be obtained on devices running
general-purpose OS and can enable detection of subtle dynamic activity changes; (2) it enables real-time exfiltration of measurements through a
streaming interface over TCP/IP to a remote monitoring agent, removing
requirement for on-device storage; (3) it enables an interactive
interface to the device implant for real-time command and control, which can be
used to dynamically query the device, perform actions, and enable a moving
target defense against adversaries; (4) it is lightweight and
scalable to support highly resource-constrained embedded devices.

\section{DIVER Framework}
\label{sec:diver}

\subsection{Threat Model}
DIVER is designed to address the threat model where an adversary
either (a) uses device vulnerabilities to gain unauthorized access and
insert/modify code to manipulate device operation; or (b) modifies network traffic or other device inputs to cause
changes to device operation.
\newcontent{
For example, in the context of VxWorks, the attacker can potentially modify any aspect of device operation,
including directly targeting the kernel through kernel modules,
impacting user-level operation through a run-time process (RTP), exploiting vulnerabilities to raise
privileges, deploying modified application-level code through unauthorized means, or manipulating
device configuration settings.
DIVER seeks to provide a vital tool for enabling deep real-time visibility into
VxWorks-based embedded devices} so as to address this
general change detection problem of flagging behavioral changes that could
be due to attacks, malfunctions, process-level changes, etc., and alerting operators on unexpected behavioral modifications. To enable such on-demand integrity verification, the defender is assumed
to be able to access the device (remotely or physically) and deploy a defensive
implant into the device RTOS using the proposed DIVER framework. Thereafter,
DIVER opens a remotely accessible port to stream measurements from the device to
enable remote monitoring and anomaly detection.
\newcontent{The use of a remote listener for measurement
  acquisition and analysis (rather than onboard analysis) is motivated by multiple reasons: (1) Embedded
   devices in CPS are typically remotely administered; hence,
  even if anomaly detection were
done on-device, it is still crucial to deliver alerts to a remote listener for remediation
actions; (2) DIVER focuses on embedded devices with stringent resource constraints
(e.g., limited/unavailable disk space to store measurements, limited memory and processing power
to analyze measurements), making it significantly more viable to perform measurement
analysis and anomaly detection at a remote machine; (3) If some level of onboard
measurement analysis/monitoring is required in some application context, on-device scripting functionalities enabled
by DIVER can be leveraged to dynamically instantiate such monitors when desired.}

\subsection{Overall Framework}
DIVER addresses an embedded device running an RTOS (specifically
VxWorks) and comprises of two components
(Figure~\ref{fig:block_diagram}):
\begin{itemize}
  \item On-device  defensive ``measurer'' implant
    deployed to the device (typically using device-specific vendor-provided
    tools such as Motorola STS software suite for Motorola ACE).
    \item Off-device ``listener'' (i.e.,  remote monitoring agent) that communicates via a network connection with the on-device component to retrieve measurements from the device and enable on-demand execution of commands on the device.
\end{itemize}
While the on-device component depends on device-specific details such as
compiler toolchains and deployment mechanisms,
its core architecture is scalable between devices and comprises of the following
primary components: a TCP server, an embedded scripting engine for run-time
configurability from the listener, a set of function callbacks exposed via
the scripting engine for use by the remote listener to invoke device operations/measurements
on demand, back-end functions that acquire measurements of device activity.
The off-device remote listener is device-agnostic and addresses
change detection and user interface.
DIVER's on-device/off-device hybrid architecture is motivated by multiple
considerations. Firstly, embedded VxWorks devices (such as Motorola ACE) do not
provide any remote access to retrieve measurements (e.g., no secure shell
access) except for device-specific HMIs which communicate using proprietary
protocols, provide limited visibility of device activity, and themselves
might be targeted by adversaries. Hence, DIVER's defensive implant
instead makes available a separate parallel communication channel
independent of vendor-provided functionalities. Secondly, the embedded devices considered are small with limited memory, storage, and processing
capabilities. Hence, instead of attempting to perform anomaly detection
algorithmic computations on the device, DIVER leverages off-device computational
capabilities at the remote listener to enable flexible algorithmic
designs for data analysis and anomaly detection while keeping the on-device
component lightweight to fit within stringent limitations of embedded
devices.  Thirdly, by decoupling on-device and off-device components, DIVER
enables device-independent off-device data analysis and interactive user
interface functionalities.

The on-device defensive implant implements a library
of measurement functions on the device that are accessible to the remote listener to execute
on-demand and retrieve data or configure for periodic execution to receive data
in streaming mode. The scripting-based automation framework embedded into the
implant enables flexible configuration of measurements and
sampling periodicities and granularities. Thus, the implant enables remote
visibility into a device that in off-the-shelf mode, does not provide any
mechanism for such visibility. Using retrieved measurements,
the remote listener can register a baseline (during training on a known-good
device) or flag anomalies (during integrity verification of a device in
the field) and more generally, detect changes in device activity over time
(same device at different points in time) or over space (across multiple devices
in the CPS
configured similarly and expected to have similar behaviors).

\subsection{Implant and Listener Architectures}
The architectures of the implant and listener components are illustrated in Figure~\ref{fig:block_diagram}.
Upon deployment, the defensive implant starts a TCP server allowing connections from remote clients. When a remote client connects, the implant initiates a query-response loop to process commands from the client. The query-response module connects to a scripting engine to perform the command-specific actions using a library of back-end functions for acquiring device-level measurements or performing device operations such as setting configuration parameters or performing I/O operations. The implant is designed to be lightweight and modular, allowing for easy addition of new measurement functions as needed. The query-response loop continues until the client disconnects or times out, after which the implant closes the connection and cleans up any remaining state from the client. The implant supports parallel connections from multiple clients to allow, for example, multiple operators connecting simultaneously through command-line or browser interfaces. After initial connection from a client, subsequent communications can use a lightweight encryption of payloads with an embedded session identifier to prevent adversaries from eavesdropping, sending unauthorized commands, or mounting replay attacks.
\newcontent{For this purpose, DIVER integrates authenticated encryption using Ascon, which is a NIST draft standard (NIST SP 800-232) providing a highly lightweight and flexible algorithm for confidentiality, integrity, and authenticity. DIVER's modular structure enables encryption to be configurable at compile-time, allowing for flexibility in deployment scenarios as well as drop-in insertions of possible future encryption algorithms. When encryption is enabled,  DIVER uses Ascon in both directions (i.e., request from remote listener, response from measurer implant). DIVER's messages also integrate timestamps and incrementing identifiers to prevent replay attacks by machine-in-the-middle attackers.}

Two variants of the scripting engine were implemented. The first is a simple text-based format allowing parameterized function calls and mathematical expression evaluations and is structured as command names with parameters to specify operations to be performed and corresponding configuration parameters (e.g., read task-level information with configurable granularity setting and specify sampling rate for streaming mode retrieval). The second is a full-featured scripting engine based on the Lua programming language, allowing for execution of more complex scripts on the device (e.g., performing a measurement or an I/O operation based on conditions such as when resource usage exceeds a specified level, setting up custom timer callbacks for operations to be performed after a specified time).

The back-end function library (details of implementation using VxWorks/device APIs discussed in next section) provides several device-level measurement functionalities such as:
\begin{itemize}
\item List of running tasks, task names and task IDs.
\item Task-level measurements of task status, program counter, stack pointer, link register, task priority, entry point, etc.; task-level activity measurements such as percentages of time in READY state, variability in program counter, etc.
\item System diagnostics (versions of VxWorks kernel and libraries, input/output configurations, uptime, system database, etc.).
\item Loaded modules and C applications with details such as file location on device, addresses of jump table and control function, hashes of initial memory segments of modules/applications, etc.
\item Readings from device diagnostics and system log mechanisms. For this purpose, the implant uses a flexible parser implemented to read binary formats used internally by device diagnostics and system log functionalities on the Motorola ACE.
\item Binary contents of ranges of memory locations.
\item Full timer tree with callback pointer locations and code hashes -- the timer tree is a hierarchical structure defining relations between the primary timer and lower-frequency timers iteratively derived from it and the callbacks registered at each level of the tree.
\end{itemize}
In addition, the implant allows actions such as
\begin{itemize}
  \item Reading/writing analog/digital inputs/outputs on the device.
  \item Reading/writing to diagnostics and system log buffers.
  \item Reading/writing to device's flash memory.
  \item Registering callback functions implemented in the scripting language to a specified timer in the timer tree to be executed periodically.
  \item Reading/writing device's time and date.
    \item Device reset.
\end{itemize}

\begin{figure*}[!h]
  \centering
  \includegraphics[width=5.5in]{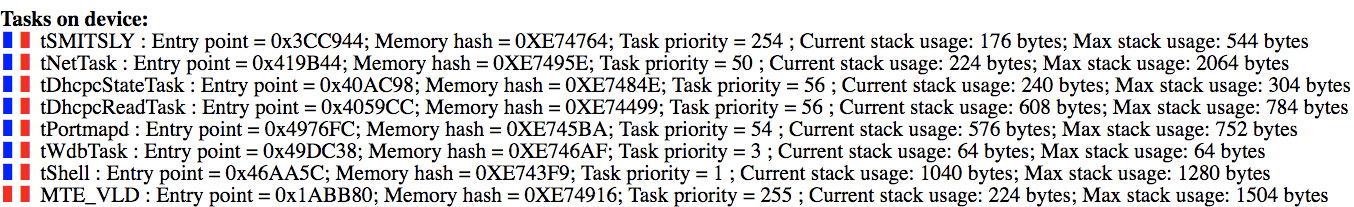}
  \hrule
  \vspace*{0.02in}
  \includegraphics[width=1.9in]{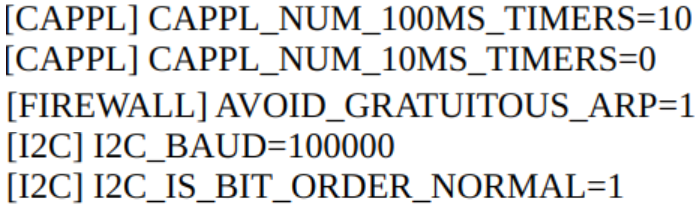}
  \ \ \
  \includegraphics[width=3in]{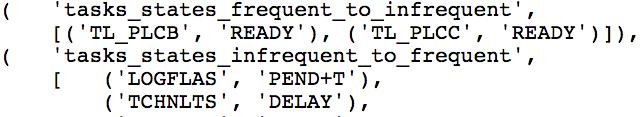}
  \hrule
  \vspace*{0.02in}
    \includegraphics[width=5.5in]{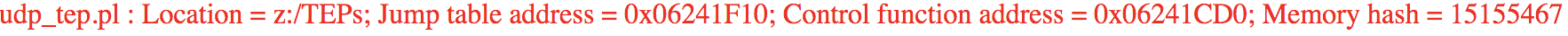}
  \hrule
  \includegraphics[width=5in]{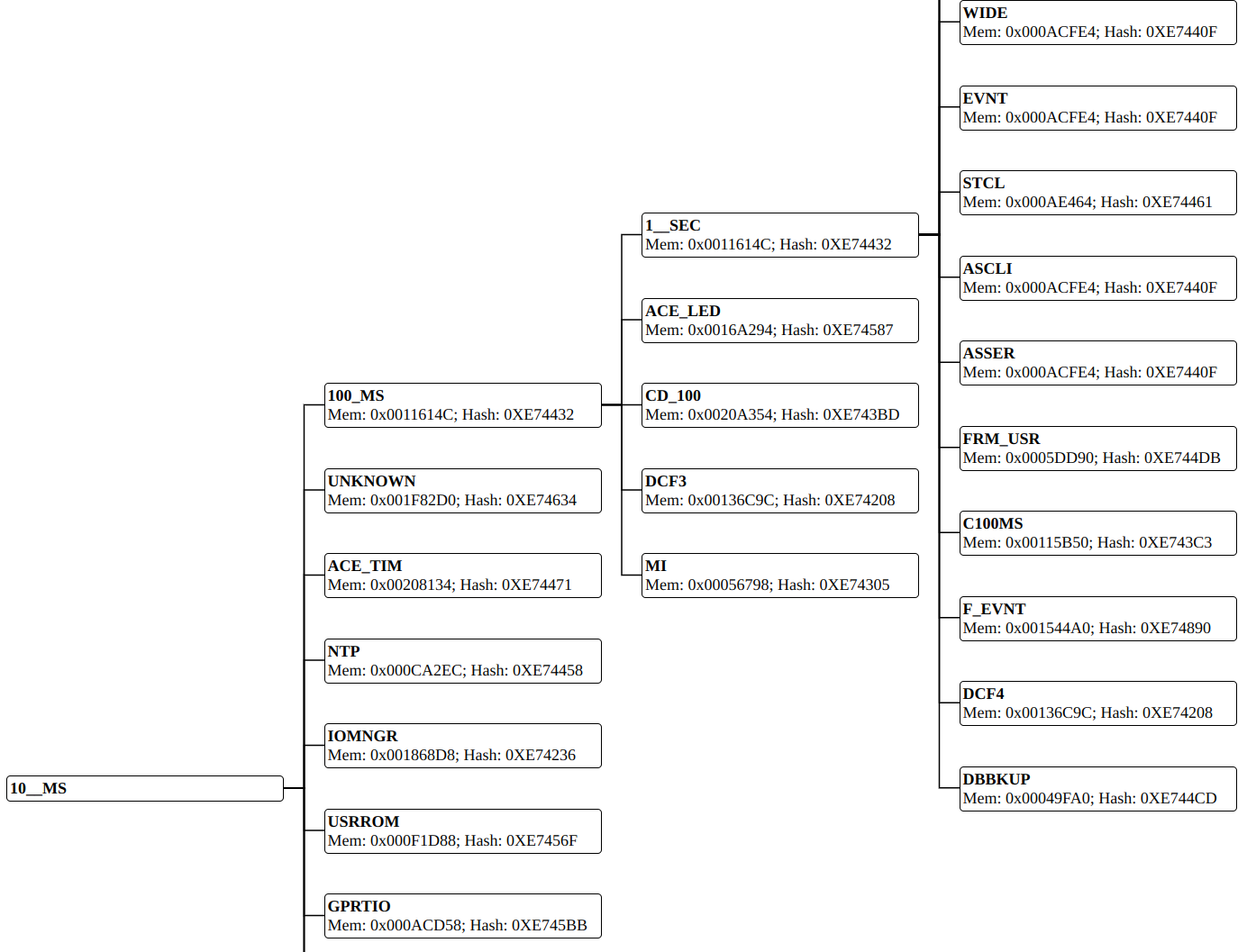}
  \hrule
   \includegraphics[width=2.5in]{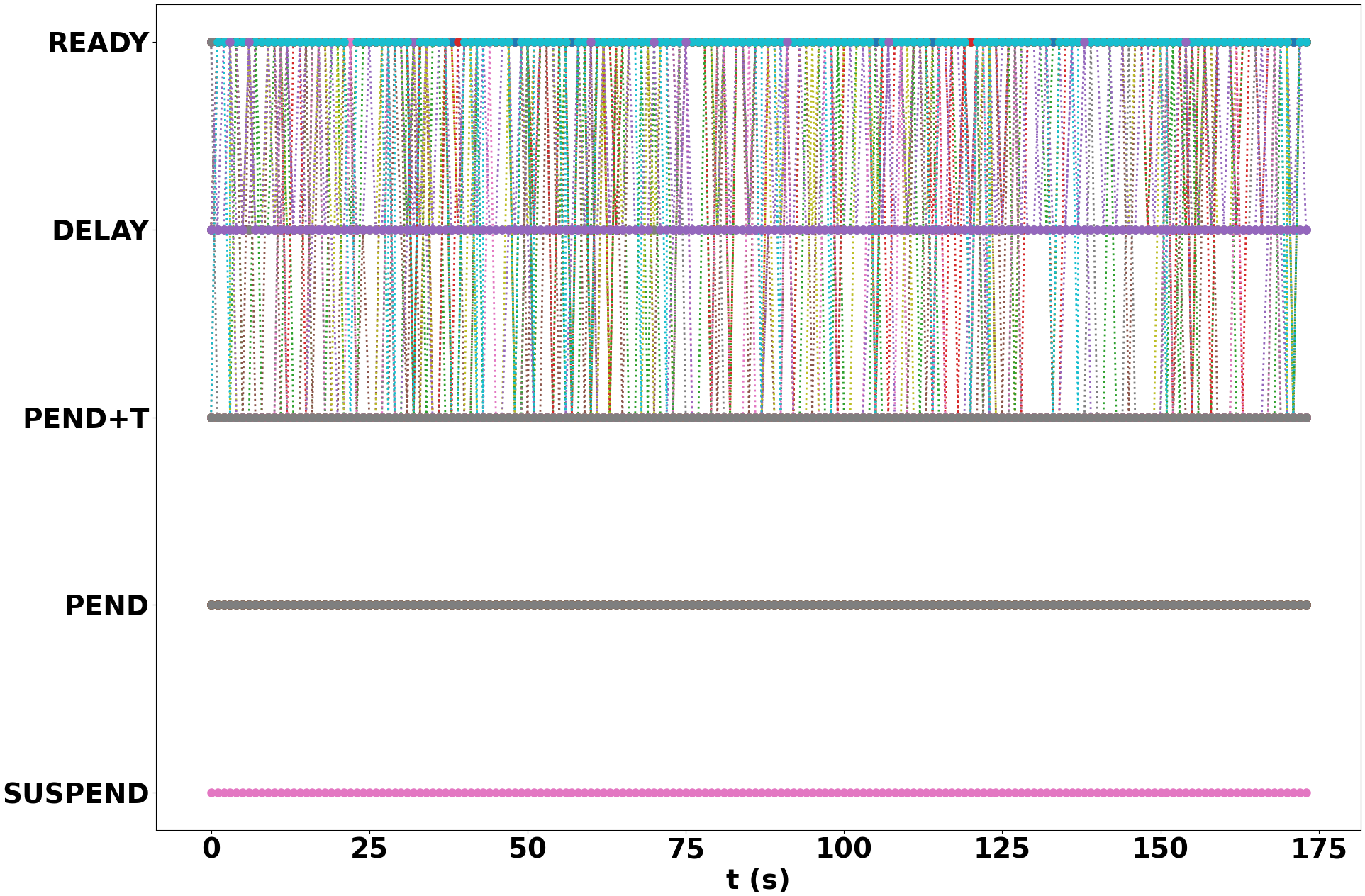}
   \includegraphics[width=2.5in]{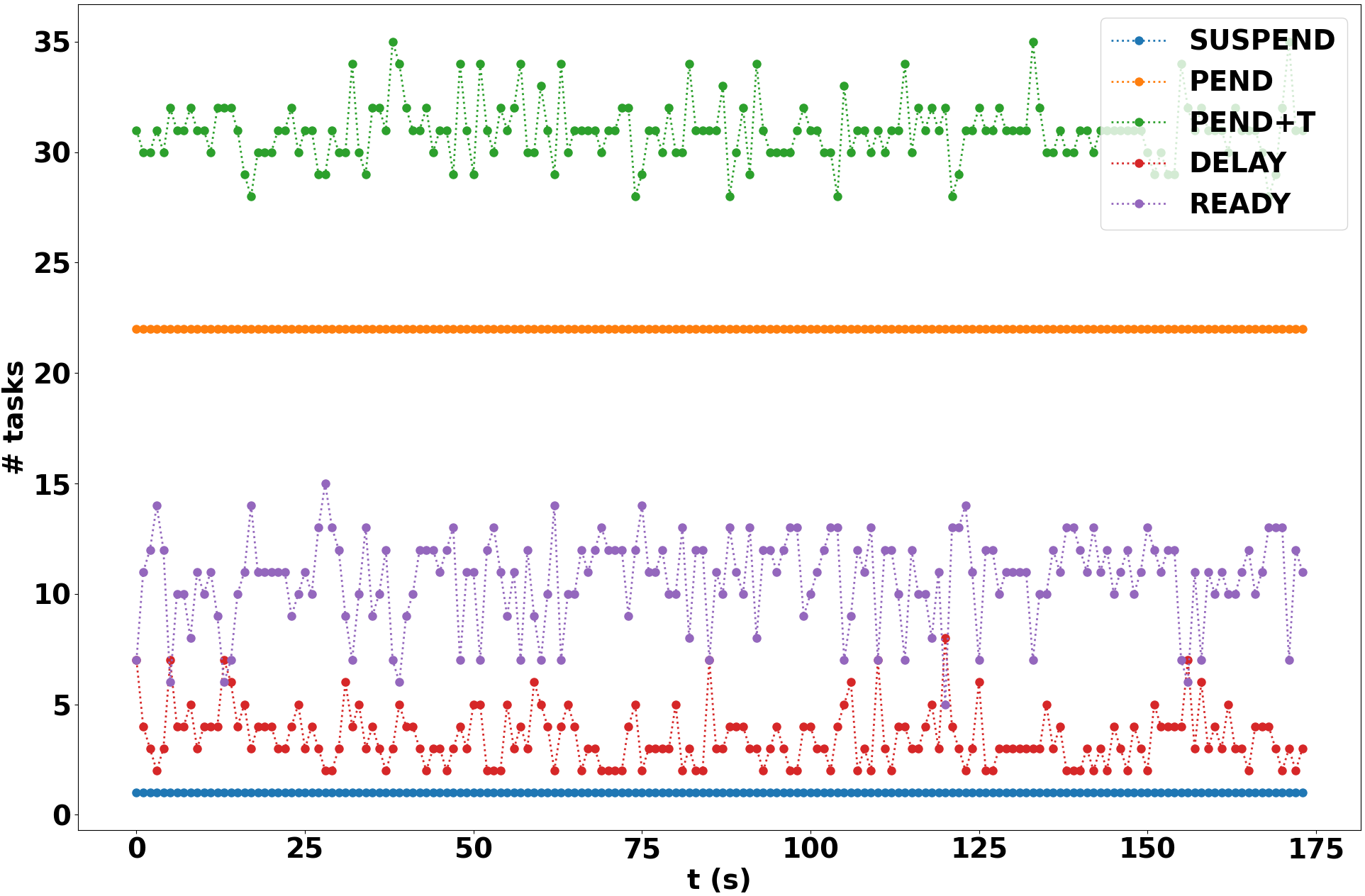}
  \vspace*{0.02in}
  \caption{Sample snippets of summaries and detected anomaly listings
    generated by remote listener using measurements exfiltrated by
    embedded implant. Top row: snippet of task activity summaries (two
    color-coded icons on left of task names indicate activity levels
    of tasks in terms of percentage of time task is in READY state and
    amount of variations seen in program counter for the task (red
    -- more active). Second row: snippets of device configuration (left)
    and task state statistics changes (right) compared to a baseline. Third row
    (in red): snippet of flagging of an unknown C application.
    Fourth row: Sample snippet of reconstructed timer tree.
    Fifth row: Sample time series of task state measurements (left) with each
    colored dotted line representing a different task and numbers of tasks
    (right) in each state (READY, PEND, etc.) at each sampling time instant.
  }
  \label{fig:sample_snippets}
\end{figure*}

The remote listener implemented in Python connects to the implant via TCP/IP and interacts with it via the query-response loop. The listener provides an interactive command-line user interface for operators to issue commands and view responses from the implant. The command-line interface is structured using the format for the simple text-based scripting engine described above. When the full-featured scripting engine is enabled, the command-line interface also provides functions for the operator to send over script snippets to be executed on the device.
The listener also provides a browser-based front-end interface, which refreshes automatically when receiving data in streaming mode and provides visualizations of retrieved data such as timer tree and color-coded visualizations of anomaly detection analysis (Figure~\ref{fig:sample_snippets}) discussed further in the next section. The browser-based interface is implemented using HTML and JavaScript with the content dynamically created by the Python-based listener.

The listener includes a data analysis module that processes retrieved measurements to build a model of device activity. This generated model can be specified to be registered as a baseline or evaluated against a previous baseline. The various types of measurements available to be exfiltrated via the implant as discussed above enable comprehensive change detection at multiple levels organized as follows:
\newcontent{\begin{itemize}
\item Checks for modifications of device configuration such as network settings, timer configurations, timeout parameters, etc.
\item Verification of enumerated kernel modules and RTPs flagging unexpected/missing elements or modifications of observed elements (e.g., changes in hashes of corresponding memory segments, changes in task priorities, etc.)
\item Verification of reconstructed timer tree (unexpected/missing timer callbacks, modifications of callback-related properties such as timing or memory segment hashes).
\item Analysis of run-time characteristics using granular measurements of task-level states, activity levels, priorities, etc., to detect statistical changes in device behavior.
\end{itemize}}
The last item above is based on the time series measurements in streaming mode, wherein the basic semantic structure for activity analysis is a time series of snapshots of running tasks with task-level observations including task states (READY, SLEEPING, etc.), task priorities, registers such as program counter and stack pointer, etc.
Variations in statistical properties of observed time series indicate fine-grained modifications in overall device activity that could result from adversarial manipulations or other changes in device operation.
For example, a task that is much more/less in READY state compared to baseline or a task that is in pending state (PEND) when it was typically in READY state in the baseline indicate anomalous changes in device operation.
The listener generates alerts/notifications in the browser-based front-end when anomalies are detected, allowing operators to take appropriate remediation actions.
\newcontent{Additionally, as noted above, DIVER's main focus is to enable deep real-time visibility into run-time operation of VxWorks-based embedded devices; DIVER-retrieved measurements and DIVER-enabled inspection of memory and filesystem contents as well as DIVER-enabled on-device scripting can be leveraged along with any binary analysis and anomaly detection algorithms in the literature.}

\section{Prototype Implementation and Experimental Studies}
\label{sec:experiments}

\newcontent{For proof-of-concept experimental studies and to demonstrate flexibility
and scalability of the framework, we deployed DIVER on two embedded devices (Motorola ACE and Raspberry Pi) with
different processor architectures and VxWorks
versions.}

Motorola ACE3600 RTU~\cite{motorola_ace} is a real-time process automation device designed
for SCADA (Supervisory Control And Data Acquisition) systems in CPS such as
power grid. ACE3600 has a 32-bit 200MHz PowerPC processor running VxWorks 5.5.
The device is modular supporting several types of digital/analog input/output (I/O) and several OT communication protocols such as Modbus, DNP3,
M-OPC, and IEC60870-5-101, as well as Motorola's own MDLC protocol.
Motorola's Windows-based ACE3600 System Tools Suite (STS) provides tools to administer/configure ACE
devices and download files/binaries to the
devices. Motorola's C Toolkit for ACE3600 RTUs with the MOSCAD
(MOtorola SCADa) APIs provides a toolchain (cross compiler, linker, etc.) for compiling C code into a
dynamically loadable module packaged into a compressed file (.plz), downloaded using STS into the device. Upon downloading,
the module is dynamically loaded and starts executing
at a predefined entry point (user\_control\_function) from which
new tasks can be spawned using VxWorks
APIs or MOSCAD APIs which are a thin layer on top of VxWorks APIs. Additionally,
functions to be periodically executed can be registered by connecting a callback pointer to
an entry in the device's timer tree (e.g., 10ms, 100ms, etc.).
To access VxWorks APIs via symbol resolution during dynamic loading, the
required VxWorks function declarations (e.g., taskIdListGet from VxWorks' taskInfo library to obtain a list of active task
IDs) are included in the C code module along with required C structure definitions (e.g., TASK\_DESC from VxWorks' taskShow library
to obtain detailed task-level information) based on VxWorks documentation.
These APIs facilitate the various implant functionalities described in the previous section such as obtaining high-granularity measurements of device activity and TCP/IP networking.

\newcontent{The second device for proof-of-concept testing is a Raspberry Pi 4,
representative of IoT applications. This device has a 64-bit ARM Cortex-A72 (ARMv8) processor and runs VxWorks 7.
While ACE runs a vendor-customized version of VxWorks,
Raspberry Pi runs standard VxWorks 7 distribution downloaded from VxWorks
(Wind River) website. Hence, while accessing underlying VxWorks functions
required additional steps for ACE as described above, the
deployment on Pi directly accesses VxWorks APIs. A major
motivation for including the Pi in experimental studies was to
study generality and scalability of DIVER across different
processor architectures (PowerPC on ACE, ARM on Pi) and
VxWorks versions (5.5 on ACE, 7 on Pi). Across the
different platforms,
DIVER uses a single codebase, intended to be portable to all
architectures supported by VxWorks. In addition to PowerPC and ARM, Intel x86-64
compatibility was also tested using VxWorks' Qemu distribution.}

\newcontent{{\bf Overhead and Benchmarks:} To evaluate DIVER's lightweight nature, we first measured its footprint and the additional overhead of its
encryption and scripting components.
Without encryption and scripting, the base measurer implant footprint was around 10.6kB on ACE and 7.3kB on Pi. Inclusion of encryption added
around 8.7kB on ACE and 7.4kB on Pi.
The simpler scripting engine that allows parameterized function calls and
expression evaluations added around 27kB on ACE and 40kB on Pi while the Lua-based full-featured scripting engine added around 250kB on ACE and 315kB on Pi.
These footprints are very small and amenable even to highly stringent resource constraints of small embedded devices in CPS and IoT applications (total footprint with encryption and lightweight scripting enabled being around 47kB on ACE and 55kB on Pi).
Recording a full task list details snapshot was measured to take around 37ms on ACE and 23ms on Pi.
Also, to measure additional overhead introduced by Ascon-based authenticated encryption, we measured time required for 10000 encryptions and decryptions of a 128-byte message.
On ACE, 10000 encryptions took 453ms and 10000 decryptions took 2076ms. On Pi, these operations required 73ms and 306ms, respectively. This corresponds to small fractions of a millisecond for encryptions/decryptions even on ACE which is slower than Pi. Hence, inclusion of Ascon authenticated encryption is viable even on small embedded devices. Furthermore, DIVER's modular structure allows encryption to be disabled at compile-time if the device is in a secure network and also allows integration of faster encryption techniques if available in the future.
}

\newcontent{
\noindent{\bf Measurement Exfiltration and Anomaly Detection:}
}
Samples of measurements retrieved using DIVER are shown in Figure~\ref{fig:sample_snippets}.
Visualizations of the sample timer tree and summary snippets in Figure~\ref{fig:sample_snippets} are from the listener's browser-based front-end.
The timer tree in the fourth row of Figure~\ref{fig:sample_snippets} shows hierarchical structure of timers on the device, with the primary timer at left and lower-frequency derived timers on right. Callbacks registered at each level of the tree are shown with corresponding memory locations and code hashes. Summary snippets in Figure~\ref{fig:sample_snippets} show various types of retrieved measurements, including task-level information and device config and example anomaly alerts.
Also, sample anomalies shown in Figure~\ref{fig:sample_snippets} indicate changes in task state statistics compared to baseline and presence of an unexpected C application along with its file location, memory addresses, and hash of the initial segment of its memory contents. The unexpected C application flagged in Figure~\ref{fig:sample_snippets} is a malware that sends a stream of UDP packets via a background malware-initiated task.
Sample time series (with 1 Hz update rate in streaming mode) of task state counts are shown in
the bottom row of Figure~\ref{fig:sample_snippets}, where each colored dotted line represents a different task and Y-axis shows different observed task states:
READY (ready to execute and waiting to be scheduled on CPU), PEND (blocked for an unavailable required resource), PEND+T (similar to PEND, but with a timeout), DELAY (sleeping for a time interval), SUSPEND (not available for execution), etc.
It is seen that there are clearly discernible temporal patterns in state transitions of tasks indicating the task behaviors.
As discussed in the previous section, the multiple types of run-time retrieved measurements enable several types of change detection.
\newcontent{In particular, in the context of the sample real-time controls application discussed further below, example attacks/anomalies tested are summarized below along with their detection using measurements retrieved by the implant:
\begin{enumerate}
    \item Insertion of a kernel module for a UDP flood attack emulating a Denial-of-Service attack.
    \item Insertion of a user-space RTP for a similar UDP flood attack.
    \item Replacement of a kernel module with another one of same name.
    \item Modified timer setting for a background callback function that logs sensor measurements.
    \item Added timer callback for timer-based data exfiltration via UDP.
    \item Modified a timer callback's memory location to replace a background callback function with an attacker-supplied one.
    \item Deletion of controller task.
    \item Disabling/suspending of controller task (overriding READY status).
    \item Modifying controller task's code with a blocking loop to make it unresponsive.
    \item Modifying priority of controller task to a lower level.
\end{enumerate}
The first two were detected as an unexpected kernel module and task, respectively. The third was detected from a mismatch in the corresponding memory hash. The next three were detected as changes in reconstructed timer tree. The remaining were flagged through mismatches against expected run-time characteristics of running tasks such as modification of task state distribution and reduced variability of program counter.
Furthermore, the deep real-time visibility enabled by DIVER into run-time operation of the embedded device provides a data source usable by any anomaly detection tools. For example, DIVER allows
on-demand exfiltration of contents of specified memory ranges or filesystem contents, which can be leveraged with any binary analysis and anomaly detection algorithms in the literature.
}

\newcontent{
\noindent{\bf Real-Time Performance Impact:}
To evaluate impact of the measurer implant on real-time performance of the embedded device, we performed experimental tests with a representative closed-loop control system application for an inverted pendulum. The pendulum is governed by the second-order dynamic model $\ddot\theta = \frac{g}{l}\sin(\theta) + \frac{1}{ml^2}u + \delta$ where $\theta$ is pendulum angle in radians, $u$ is control input torque, $g=9.81 m/s^2$ is acceleration due to gravity,  $m$ and $l$ are mass and length of the pendulum, respectively, and $\delta$ is noise (e.g., torque disturbance). For testing, we set $m=2$ kg, $l=0.5$ m, and initial $\theta=0.5$ rad. We model $\delta$ as zero-mean Gaussian with standard deviation 1. The control law designed as $u=-k_1\theta-k_2\dot\theta$ with $k_1=12.5$ and $k_2=2.25$ is implemented as a separate controller task in the VxWorks RTOS on the embedded device. The pendulum dynamic system simulation on a computer and the on-device controller communicate using UDP messages with $\theta$ and $\dot\theta$ transmitted from the simulator at 100 Hz to the controller, which responds with the control commands. The real-time performance of the closed-loop system is shown in Figure~\ref{fig:sim_measurements} where first two rows are with the controller implemented on ACE and last two rows are with the controller on Pi. In the second and fourth rows, the measurer implant is deployed and streams task state measurements at 10 Hz to the listener. A small increase in jitter is seen in control response times with the measurer, but this does not impact the closed-loop performance which still remains stable. With the ACE, control response times are observed to be 3.46 ms without and 3.48 ms with the measurer. Corresponding times with Pi are 0.4 ms and 0.44 ms, respectively. These observations underscore DIVER's lightweight nature and real-time performance for run-time visibility into the embedded device. Furthermore, while the experimental tests above use the same ethernet connection for both UDP-based controller implementation and measurer, a real-world deployment with physical digital/analog I/O in a CPS would show even lower impact by being independent of the measurer's network communications.

\begin{figure*}[!h]
\centering
\includegraphics[width=6in]{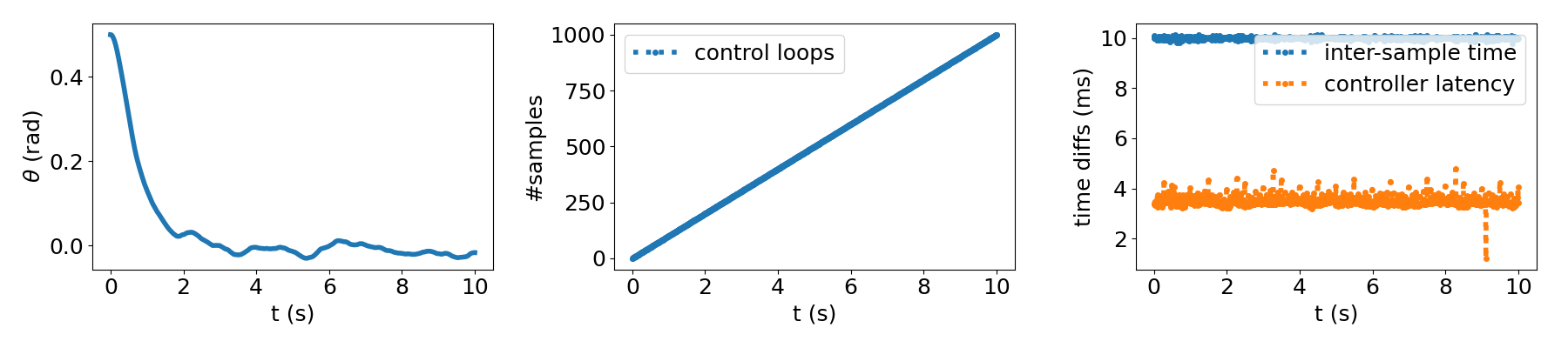}
\includegraphics[width=6in]{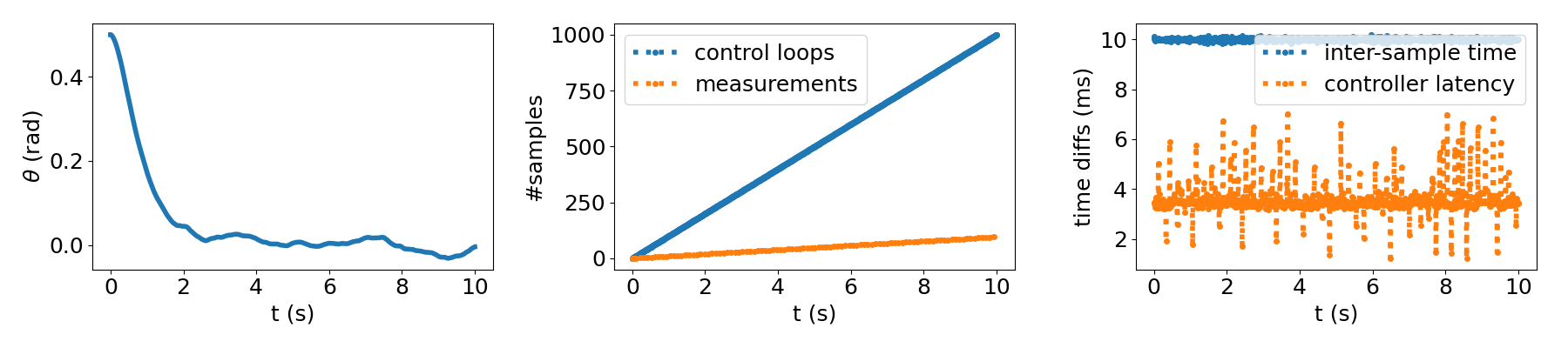}
\includegraphics[width=6in]{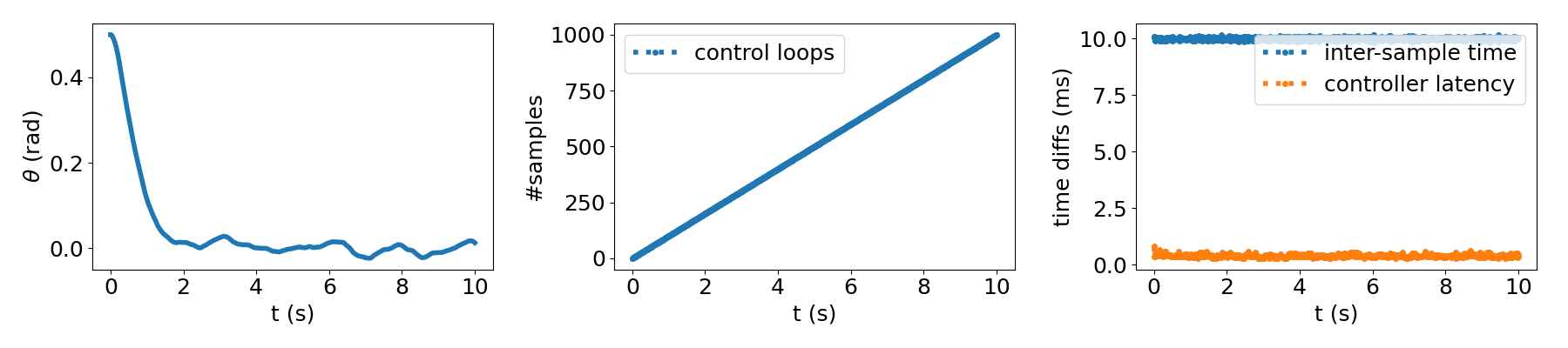}
\includegraphics[width=6in]{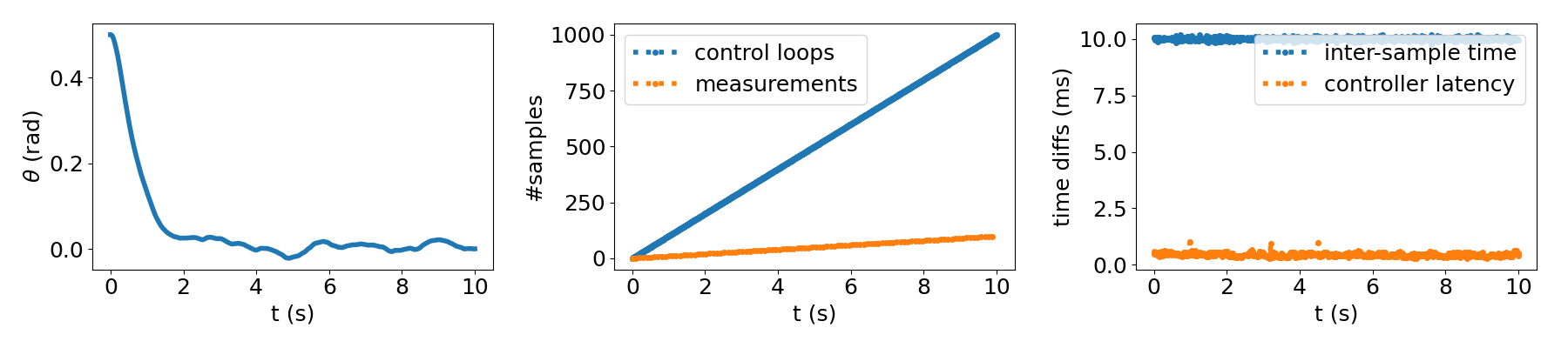}
\caption{Real-time performance in a closed-loop inverted pendulum system with controller implemented on embedded device (first two rows: Motorola ACE, last two rows: Raspberry Pi). The first and third rows are without the measurer implant. The second and fourth rows are with the implant streaming measurements. Each row shows: Left -- pendulum angle; Middle -- number of samples (control loops, measurement samples when measurer active); Right -- time differences between successive control command samples, latency from sensor data transmission to control command receipt.}
\label{fig:sim_measurements}
\end{figure*}

}

\section{Discussion and Conclusions}
\label{sec:conclusions}
DIVER focuses on VxWorks-based lightweight embedded devices due to their prevalence in CPS such as power grid and industrial control systems and due to the unique challenges posed by such devices with stringent resource constraints and on-device limitations precluding traditional monitoring methods.
By deep integration into VxWorks RTOS, DIVER enables deep real-time visibility to flag anomalies in device operation.
DIVER can also indirectly enable detection of attacks on HMIs by enabling flagging mismatches between on-device measurements and HMIs' view of device operations.
DIVER's real-time visibility into device operation can also be used to extract salient properties of device activity, e.g., to identify tasks with most activity or using most resources (i.e., analogous to a tool such as \verb~htop~ on Linux).
DIVER is designed to be adaptable to different embedded devices and to be extensible to additional types of measurements and anomaly detection techniques. Also, DIVER's modular on-device/off-device structure and real-time visibility of device activity and exfiltration of time series measurements can benefit third-party monitoring and event detection systems to assist in remote integrity verification.
Future work will focus on enhancing the framework to support additional types of measurements and anomaly detection techniques, and evaluating on broader range of embedded devices.

\bibliographystyle{IEEEtran}
\bibliography{refs}

\end{document}